# Textured Superconductivity in the Presence of a Coexisting Order: Ce115s and Other Heavy-Fermion Compounds

Tuson Park[a,b], Xin Lu[b], Han-Oh Lee[b,1] and J. D. Thompson[b*]

[a]Department of Physics, Sungkyunkwan University, Suwon, 440-746, Korea

[b]Los Alamos National Laboratory, MS K764, Los Alamos, NM 87545, USA

email addresses:

Tuson Park: tp8701@skku.edu

Xin Lu: xinlu@lanl.gov

Han-Oh Lee: peartree77@gmail.com

J. D. Thompson: jdt@lanl.gov

*Corresponding author: J. D. Thompson; tel: (505) 667-6416; fax: (505) 665-7652

[1]Present address: Department of Applied Physics and Geballe Laboratory for Advanced Materials, Stanford University, Stanford, CA 94305 USA

## Abstract

Superconductivity in strongly correlated electron systems frequently emerges in proximity to another broken symmetry. In heavy-electron superconductors, the nearby ordered state most commonly is magnetism, and the so-called Ce115 heavy-electron compounds have been particularly instructive for revealing new relationships between magnetism and superconductivity. From measurements of the resistive and bulk transitions to superconductivity in these materials, we find that the resistive transition appears at a temperature considerably higher than the bulk transition when superconductivity and magnetic order coexist, but this temperature difference disappears in the absence of long-range magnetic order. Further, in the pressure-temperature region of coexistence in $CeRhIn_5$, a new anisotropy in the resistive transition develops even though the tetragonal crystal structure apparently remains unchanged, implying a form of textured superconductivity. We suggest that this texture may be a generic response to coexisting order in these and other heavy-fermion superconductors.

Key words: $CeRhIn_5$, textured superconductivity, antiferromagnetism, coexisting orders, nematicity

## 1. Introduction

Evidence for electronic nematic or smectic phases that break rotational or translational as well as rotational crystal symmetry in the cuprates is well-established by a variety of techniques, including electrical transport [1], neutron [2,3] and x-ray [4] scattering, scanning tunneling spectroscopy [5-7], and Nernst effect [8] measurements. Though experiments have established the existence of these symmetry-breaking phases, their physical origin and relationship to cuprate superconductivity remain unclear.[9] This also is the case for the iron-arsenide superconductors in which experiments find spontaneous broken rotational symmetry in underdoped $Ca(Fe_{1-x}Co_x)_2As_2$ [10] and $Ba(Fe_{1-x}Co_x)_2As_2$ [11]. The discovery of a high nematic susceptibility in this new family of materials suggests generality of the phenomenon, but the influence of crystal-structure distortions from simple tetragonal symmetry, of twinning and of the inevitable presence of chemical and associated structural disorder has made it difficult to isolate intrinsic contributions to broken symmetry phases in both cuprate and iron-arsenide systems.[9,12] Each of these complications contributes to the complex real-space electronic texture that is characteristic of a nematic phase. The strongly correlated heavy-fermion systems, however, offer a route to exploring the emergence of spontaneous symmetry breaking and electronic texture without these extrinsic contributions. Like the cuprates and iron-arsenides, superconductivity in heavy-fermion materials develops in proximity to long-range magnetic order [13]; however, the superconductivity either is present in pristine crystals under ambient conditions or is accessed by a 'clean' tuning parameter, pressure, without the need for chemical doping. The family of so-called Ce115 heavy-fermion materials, $CeRhIn_5$, $CeCoIn_5$ and $CeIrIn_5$,[14] are additionally similar to the curpates and iron arsenides in that they crystallize in a tetragonal structure, which is built from a magnetic layer of $CeIn_3$ units analogous to the $CuO_2$ planes in cuprates and Fe-As planes in the arsenides. In spite of these similarities, the possibility of a textured electronic phase in the Ce115s has been considered only recently.[15]

The temperature-pressure phase diagram of $CeRhIn_5$, shown in Fig. 1(a), is typical of several heavy-fermion systems in which pressure induces superconductivity from an antiferromagnetic state at ambient pressure. In this material, there is a range of pressures below P1(=1.75 GPa) where bulk d-wave superconductivity coexists microscopically [16-19] with large-moment, incommensurate, **Q**= (½, ½, 0.297), antiferromagnetic order.[20] In the coexistence regime, the electrical resistivity drops to an immeasurably small value (typically less than 1 to 5 nΩcm) at a temperature notably higher than the bulk transition determined by specific heat,[18,19] but as illustrated in this figure, once evidence for antiferromagnetic order disappears at pressures P > P1 the bulk and resistive transition temperatures coincide. It seems highly unlikely that these observations are due to chemical or structural disorder because at atmospheric pressure the residual resistivity ratio ρ(300K)/ρ(T →0) is over 700 in these crystals, with the same response in crystals with lower resistivity ratio of about 200 [21], and a nearly hydrostatic pressure environment should not remove any potential disorder. Instead of poor sample quality, the evolution of resistive and bulk superconducting transition temperatures suggests the possible presence of some form of electronic texture at pressures below P1.

A theoretically proposed relationship [22] between superconductivity, nematic and smectic phases in the cuprates is reproduced in Fig. 1(b). In this model, the energy ħω of transverse, zero-point fluctuations of composite spin and/or charge density waves induces the change of an electronic solid to an electronic smectic liquid and eventually to an electronic nematic, concurrently enhancing the propensity of carriers to form singlet Cooper pairs. The T=0 boundaries of smectic, nematic and superconducting phases define quantum-critical points.[22] Pressure applied to CeRhIn$_5$ also induces a quantum-critical state, with a line of quantum-phase transitions revealed inside the superconducting dome when a magnetic field is applied.[18,19] Zero-point fluctuations emerging from this criticality favor singlet (d-wave) Cooper pairing [18], and in this case the fluctuations are dominantly magnetic in character but also may be accompanied by fluctuations in valence [23]. Though a direct analogy between phase diagrams in Figs. 1(a) and (b) is unlikely, their similarity is suggestive and motivates, in part, an exploration for evidence of electronic texture in CeRhIn$_5$. In particular, the theoretical phase boundaries delineating the smectic and nematic phases extend into the superconducting state, raising the possibility that a signature of their real-space texture might be reflected in the superconducting transition. As will be reviewed, a new anisotropy develops in the resistively measured transition in CeRhIn$_5$ that is consistent with the formation of textured superconducting lamellae preferentially oriented in {110} planes. The absence of this anisotropy at pressures P > P1 suggests that the textured superconductivity is associated with coexisting antiferromagnetic order. Though our study of CeRhIn$_5$ has revealed this texture, it could be a consequence of a generic order coexisting with superconductivity in strongly correlated systems, and, indeed, suggestive evidence is found in other heavy-fermion superconductors with coexisting strong commensurate antiferromagnetism, with weak incommensurate spin density order, and with possibly some form of order yet to be identified.

## 2. Material and methods

Crystals of CeRhIn$_5$, which form in the tetragonal HoCoGa$_5$ structure, were grown from an In flux and screened by SQUID magnetometry with an applied field of 10 Oe to ensure the absence of free In. This is the case as well for family members CeIrIn$_5$ and Cd-doped CeCoIn$_5$, which will be discussed. To increase the sensitivity of anisotropic resistivity measurements, crystals were polished into bar shapes with long dimensions along [100], [001] and [110] directions, and a conventional four-probe technique was used to measure the resistivity with current flow in the long dimension. The specific heat of CeRhIn$_5$ was determined by a semi-quantitative *ac* calorimetry technique.[24] Pressures to 2.8 GPa were generated in a NiCrAl/BeCu hybrid-type clamp cell using a silicon fluid to produce a nearly hydrostatic pressure environment. The pressure at low temperatures was measured by the suppression of the superconducting transition of Sn.

## 3. Results and discussion

*3.1 CeRhIn$_5$*

Typical specific heat C and in-plane electrical resistivity $\rho_{ab}$ data for CeRhIn$_5$ are plotted in Figs. 2(a) and (b) at a representative pressure below and above P1. Data such as these are reproducible in all of several crystals that have been studied and are used to construct the phase diagram in Fig. 1(a). Below P1, Fig. 2(a), there are well-defined anomalies in specific heat near 2.2 K and 1.2 K that signal bulk antiferromagnetic ($T_N$) and superconducting ($T_c$) transitions, respectively. On the other hand, the in-plane resistivity drops rapidly but incompletely toward zero at an intermediate temperature and is followed by a long tail that becomes immeasurably small at the bulk $T_c$. With increasing pressure, the difference between the temperature at which

$\rho_{ab}$ initially drops toward zero and $T_c$ decreases and becomes less than a few milliKelvin above P1 where antiferromagnetic order is absent. At this higher pressure, Fig. 2(b), the resistive transition is very sharp and coincides with $T_c$ determined by specific heat. This evolution of $\rho_{ab}$ and its relationship to $T_c$ indicate clearly a dependence on the presence of antiferromagnetic order. Further, when the measuring current is varied by two orders of magnitude, from 0.1 to 10 mA, and hence the input power by four orders, there is negligible change in shape of the resistive transition, and we conclude that the effects shown in Fig. 2 are intrinsic to these very pure single crystals. The in-plane resistive transition is qualitatively different from the transition measured with current flow along the c-axis, as shown in Fig. 3(a). Though the pressures for which data are plotted here are marginally different than those in Fig. 2 and very slightly different for current in the a-b plane and along the c-axis, these curves illustrate the essential points that the c-axis resistive transition sharply approaches zero resistance even for P < P1 and that anisotropy in the resistive transition disappears above P1. (We note that there is little anisotropy in the low-temperature normal-state resistivity at pressure below P1 or just above P1. For example, at 3 K $\rho_{ab}/\rho_c$ =1.2 and 1.1 at 1.48 GPa (<P1) and 1.81 GPa (>P1), respectively.[23]) The physical interpretation implied by these results is that, in the presence of long-range antiferromagnetic order, a connected path of superconductivity forms above the bulk $T_c$ and that either separately or through a connected network extends spatially from one side of the crystal to the other along the c-axis but are not continuous across directions perpendicular to the c-axis. As temperature is lowered further toward $T_c$, the lamellar network eventually creates a continuous superconducting path in the a-b plane. This lamellar texture appears analogous to stripes in the cuprates, and, indeed, Fig. 3(b) shows that the in-plane resistive transition in $La_{1.875}Ba_{.125}CuO_4$ [25] resembles that plotted in Fig. 3(a) for $CeRhIn_5$ at 1.61 GPa. At this hole doping in the cuprate, the initial order of magnitude drop in resistivity at ~40 K appears nearly coincident with spin order in stripes. [25] We cannot say from these experiments on $CeRhIn_5$, however, if the lamellae are static or dynamic or what symmetry they might break.

To explore possible in-plane symmetry breaking, we have measured the resistive transition for current flow along [100] and [110], with results given in Fig. 4. For these studies one crystal was cut into two bar-shaped pieces whose long dimension was determined by x-ray diffraction to be along [100] and [110]. The resistive transition along [100], $\rho_{[100]}$, is very similar to that of $\rho_{ab}$ plotted in Fig. 2 (a) at a slightly higher pressure and is consistent with our estimate of the orientation of current flow in the experiment at 1.6 GPa. The significant results illustrated in Fig. 4 are that the long resistive tail to zero resistance in $\rho_{[100]}$ is absent in $\rho_{[110]}$ and that there is a pronounced in-plane anisotropy in the resistive transition. Though not shown here, experiments at lower and higher pressures find [15] that the in-plane anisotropy, characterized by the midpoint of the transition, grows at pressures lower than 1.45 GPa, decreases at higher pressures and is undetectably small for P > P1. This anisotropy, present only when incommensurate antiferromagnetism coexists with superconductivity, appears to break the four-fold rotational symmetry of the crystal lattice. Though measurements of the crystal structure have not been made explicitly at these temperatures and pressures, there is indirect evidence that the structure is unchanged or at least, if changed, is undetectably small: (1) single-crystal magnetic neutron-diffraction experiments at these pressures and temperatures do not reveal a change in crystal structure; [20] (2) nuclear quadrupole-resonance spectra, which are sensitive to the local crystalline environment of the NQR nucleus, evolve smoothly without evidence for a structural change; [16] and, (3) deHaas-vanAlphen frequencies, and hence presumably the electronic and crystal structures, are unchanged for pressures below P1.[26] Detailed measurements of $\rho_{[100]}$ and $\rho_{[110]}$ find that the anisotropy in their transitions is robust against a two-order of magnitude change in measuring current, again indicating that it is intrinsic. Anisotropy in the resistively determined transition in CeRhIn$_5$ suggests the presence of an intrinsic region intermediate to the onset of pair formation and the bulk $T_c$ in which superconductivity is textured in real space due to coexisting antiferromagnetic order.

*3.2 Other Ce115s*

CeRhIn$_5$ is not alone in showing a difference between bulk and resistive transitions to a d-wave superconducting state. Unlike CeRhIn$_5$, CeCoIn$_5$ and CeIrIn$_5$ are superconducting at atmospheric pressure; however, commensurate antiferromagnetic order can be induced in CeCoIn$_5$ by replacing a small number of In atoms with Cd.[27,28] For a range of Cd concentrations, ~ 0.6 to 1.2 atomic percent Cd per In atom, antiferromagnetic order is established before superconductivity at a lower temperature.[29] Neutron diffraction [27,28] and NMR [30] experiments show that both orders coexist microscopically. Applying pressure to these Cd-doped samples suppresses the antiferromagnetism leaving only bulk superconductivity [29], similar to the case of CeRhIn$_5$. As shown in Fig. 5(a) for CeCo(In$_{.99}$Cd$_{.01}$)$_5$, when the orders coexist the transition to an immeasurably small resistance state occurs at a temperature approximately midway between the Néel temperature and bulk $T_c$ determined by specific heat. At a pressure sufficiently high to suppress long range magnetic order, for example at 1.5 GPa shown in Fig. 5(b), the in-plane resistive and bulk superconducting transitions coincide as they do in CeRhIn$_5$ at P > P1 and in undoped CeCoIn$_5$ at atmospheric pressure. We have not explored evidence for anisotropy in the resistive transition of Cd-doped CeCoIn$_5$, but the results of Fig. 5 indicate the establishment of superconducting filaments or a connected network of filaments at a temperature well above the bulk $T_c$. An important conclusion from Fig. 5 is that a difference between bulk and resistive $T_c$'s appears irrespective of whether the coexisting antiferromagnetic order is commensurate (½, ½, ½) [27], as in this case, or incommensurate (½, ½, δ) [31], as in CeRhIn$_5$.

Though forming in the same crystal structure as CeRhIn$_5$ and CeCoIn$_5$ and as crystals with comparably large resistivity ratios, CeIrIn$_5$ is unusual: its bulk $T_c$ is 0.4 K but the resistive transition occurs near 1 K or higher (see Fig. 6(a)).[32] This difference is found in all crystals, irrespective of their source, and given results on CeRhIn$_5$ and Cd-doped CeCoIn$_5$ might not be surprising if there were magnetic order in CeIrIn$_5$ above $T_c$; however, no bulk phase transition has been found other than superconductivity. Nevertheless, as shown in Fig. 6(b), $T_c$ is a non-monotonic function of pressure, reaching a maximum near 3 GPa [33] where the nearly pressure independent in-plane resistive midpoint transition [34] approaches the bulk $T_c$. The dome-like shape of $T_c(P)$ suggests a pressure-dependent competition between superconductivity and some other phase that could be responsible for results plotted in Fig. 6(a). Indeed, careful Hall effect and magnetoresistance studies of CeIrIn$_5$ find evidence for a precursor state, similar to a pseudogap in the cuprates, that develops near 2 K in the limit of zero magnetic field.[35] We have not determined if the resistive transition breaks rotational symmetry in the basal plane, but, interestingly, the resistivity drops to an immeasurably small value at a notably higher temperature for current along [100] than for current parallel to the c-axis. This anisotropy is opposite that found in the coexistence phase of CeRhIn$_5$ and indicates that a connected path of textured superconductivity forms preferentially in the basal plane. If the precursor state were to have the character of the cuprates' pseudogap, then the so-far unidentified competing order might be analogous to an electronic nematic phase in which superconducting pairs form initially along one- or two-dimensional spin/charge networks that eventually lead to bulk superconductivity at a lower temperature.

*3.3 CeCu$_2$Si$_2$ and URu$_2$Si$_2$*

The possibility that textured superconductivity is not a peculiar property of the Ce115s is indicated from specific heat and resistivity measurements on the heavy-fermion compounds CeCu$_2$Si$_2$ and URu$_2$Si$_2$. Metallurgically, CeCu$_2$Si$_2$ is much more complex than the Ce115s and can form crystals that are only superconducting (S-type), only antiferromagnetic (A-type) or both superconducting and magnetic (A/S-type). In A/S-type crystals, the resistive transition to a superconducting state develops at a temperature nearly twice the bulk $T_c$ as illustrated in Fig. 7(a).[36] In this case, the order competing with superconductivity is a weak incommensurate spin density wave.

Tetragonal URu$_2$Si$_2$, structurally equivalent to CeCu$_2$Si$_2$, has received much attention because bulk superconductivity below 1.2 K coexists with a 'hidden order' that develops at 17.5 K.[37] Unlike CeIrIn$_5$, however, the hidden order manifests itself in a pronounced second-order phase transition. As discussed in this issue [38], torque magnetometry reveals an in-plane anisotropy that breaks four-fold symmetry of the crystal lattice in the hidden order phase and that is consistent with the emergence of an electronic nematic state that leaves the lattice translationally invariant [39]. These experiments are perhaps the strongest evidence for electronic nematicity in a heavy-fermion compound. Like other examples cited above, there also is a difference between the resistive and bulk $T_c$'s in URu$_2$Si$_2$ (Fig. 7(b)). [40] Whether this difference is due to electronic nematicity or to some other aspect of the hidden order in URu$_2$Si$_2$ remains to be determined, but the apparent ubiquity of such differences in several heavy-fermion compounds suggests that real-space texture in their superconductivity maybe a generic response of these correlated electron systems when there is a coexisting order.

**4. Conclusions**

The appearance of symmetry-breaking anisotropies in electrical transport has been useful for identifying evidence for real-space electronic texture in cuprates and iron-arsenides, which subsequently was verified directly by imaging the structure through scanning tunneling spectroscopy. We have used resistivity measurements to uncover intrinsic texture in the d-wave superconductivity of CeRhIn$_5$ when it coexists with large-moment incommensurate order. This texture is accompanied by a pronounced difference between bulk and resistive transition temperatures, which also is found in other members of the Ce115 family and in CeCu$_2$Si$_2$ when their superconductivity coexists with commensurate antiferromagnetic order, with some other competing state that is pseudogap-like or with a weak incommensurate spin density. URu$_2$Si$_2$, so far the clearest example of electronic nematicity in heavy-fermions, exhibits a similar disparity between superconducting bulk and resistive transitions. We do not understand the origin of textured superconductivity, but it appears to be a consequence of some coexisting state which itself may determine anisotropy of the texture, eg., through preferential orientation of domain boundaries. An important issue for further study is to clarify the extent to which these features in heavy-fermion materials are a consequence of the electronic smectic or nematic states that have posed such interesting problems in the cuprates.

**Acknowledgement**

We thank E. D. Bauer and Z. Fisk for providing Ce115 samples and S. A. Trugman, I. Martin and F. Ronning for helpful discussions. Work at Los Alamos was performed under the auspices of the U.S. Department of Energy, Office of Basic Energy Sciences, Division of Materials Sciences and Engineering. TP acknowledges support by the National Research Foundation of Korea, NRF No. 2010-002672.

**Figure captions**

Fig. 1 (color online) Experimental phase diagram for $CeRhIn_5$ and theoretically predicted diagram for nematic and smectic order in a doped Mott antiferromagnet. (a) Temperature versus pressure diagram for $CeRhIn_5$. Triangles denote the Néel boundary (AFM); stars and open circles represent the resistive mid-point and bulk superconducting transitions (SC), respectively. P1 defines the pressure above which magnetic order is absent in zero applied magnetic field. A line of field-induced quantum criticality appears between P1 and the pressure where $T_N(P)$ extrapolates to T=0 [18]. (b) Evolution of crystalline, smectic, nematic and superconducting phases as a function of the energy of transverse, zero-point fluctuations of stripes. Adopted from [22].

Fig. 2 (color online) Specific heat divided by temperature C/T and in-plane electrical resistivity $\rho_{ab}$ of $CeRhIn_5$ as a function of temperature for a pressure less than and greater than P1, plotted in panels (a) and (b), respectively. Filled circles denote C/T (left ordinate) and filled triangles show resistivity (right ordinate).

Fig. 3 (color online) Resistive transitions in $CeRhIn_5$ and $La_{1.875}Ba_{.125}CuO_4$. (a) In-plane $\rho_{ab}$ (solid symbols) and c-axis $\rho_c$ (open symbols) resistivity on a logarithmic scale as a function of temperature for $CeRhIn_5$ at a pressure less than P1 (up triangles) and at a pressure greater than P1 (down triangles). The in-plane and c-axis resistivity was measured on separate crystals from the same batch and was reproduced in separate measurements on different crystals. (b) In-plane resistive transition of $La_{1.875}Ba_{.125}CuO_4$. The order of magnitude drop in $\rho_{ab}$ occurs at a temperature much above the bulk superconducting transition temperature indicated by an arrow. After Ref. 25.

Fig. 4 (color online) Temperature dependence of the electrical resistivity of $CeRhIn_5$ at a pressure less than P1 with current flow along [100] (filled circles) and along [110] (open circles).

Fig. 5 (color online) Specific heat divided by temperature C/T (filled circles, left ordinate) and in-plane electrical resistivity $\rho_{ab}$ (filled triangles, right ordinate) of CeCoIn$_5$ doped with 1% Cd. (a) At atmospheric pressure, Néel order sets in near 2.8 K, well above the bulk T$_c$ just above 1 K. (b) Applying 1.5-GPa pressure to this sample suppresses magnetic order, leaving only superconductivity near 2.3 K.

Fig. 6 (color online) Experimental results for CeIrIn$_5$. (a) Temperature-dependent specific heat divided by temperature C/T (filled circles, left ordinate) and normalized resistivity for current flow along [001] (solid symbols) and along [100] (open symbols) at atmospheric pressure. At 1.6 K, $\rho_{[100]}$ = 1.8 μΩcm and $\rho_{[001]}$ = 3.9 μΩcm. (b) Pressure dependence of the bulk (circles) and mid-point of the in-plane resistive (triangles) transitions to superconductivity. Data for the bulk T$_c$ are taken from Ref. 33 and those for the resistive transition from Ref. 34.

Fig. 7 (color online) Specific heat divided by temperature C/T (solid circles) and electrical resistivity ρ (solid triangles) versus temperature for CeCu$_2$Si$_2$ and URu$_2$Si$_2$. (a) The transition to zero resistance for an A/S-type crystal of CeCu$_2$Si$_2$ appears just below the Néel temperature at 0.7 K; whereas, the bulk T$_c$ is near 0.4 K. Adopted from [36]. (b) Superconducting transitions determined by specific heat and resistivity on a single crystal of URu$_2$Si$_2$. Data are adopted from [40].

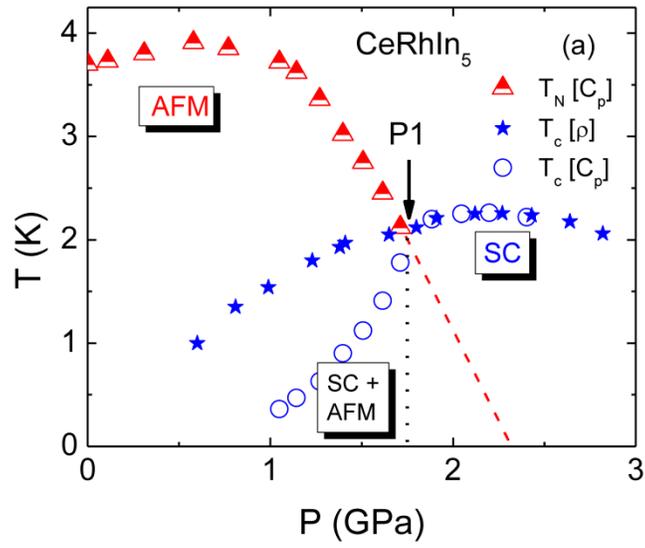

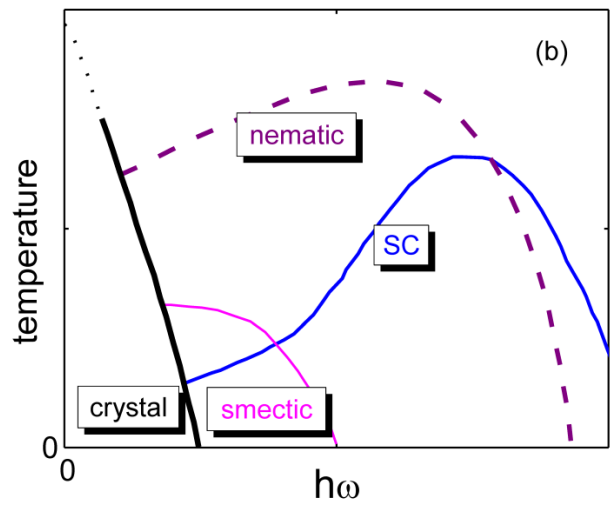

Figure 1

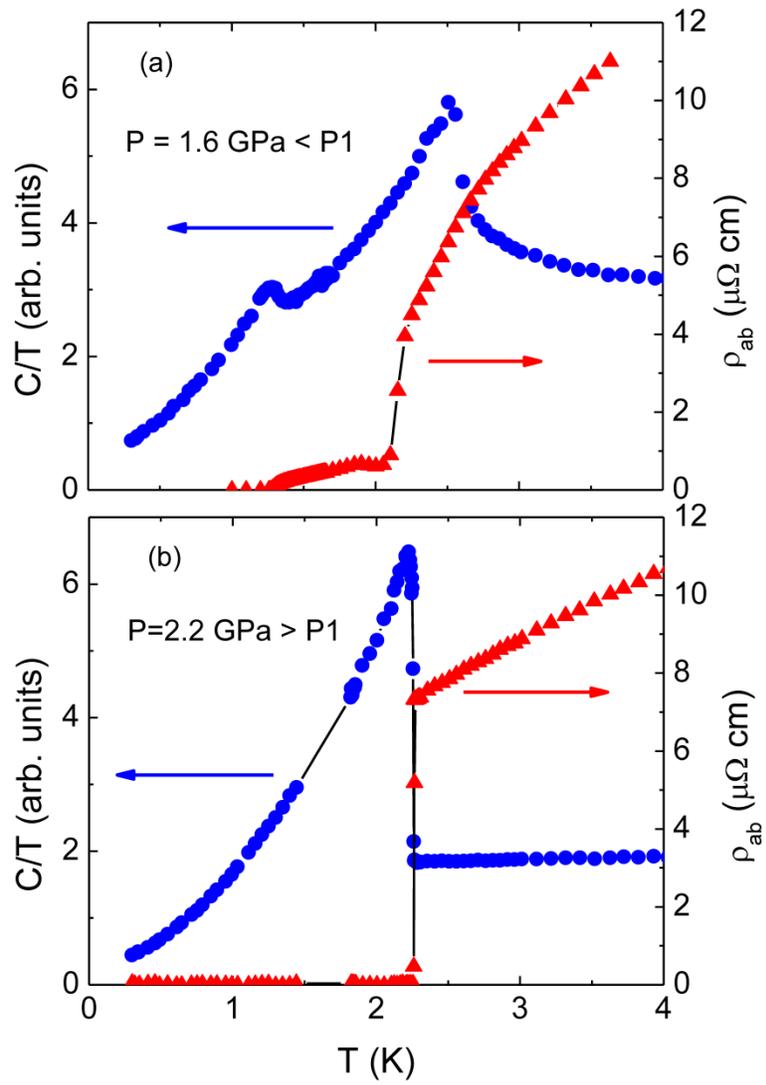

Figure 2

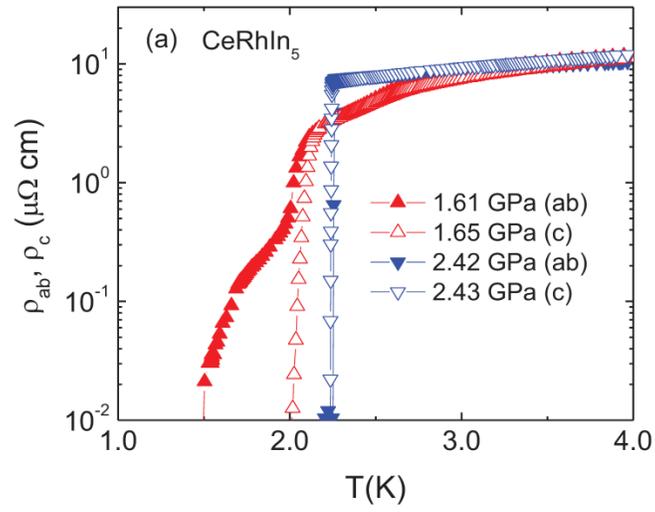

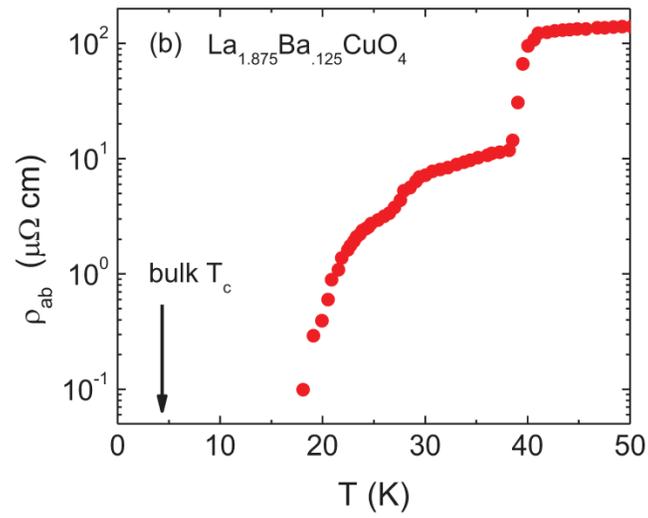

Figure 3

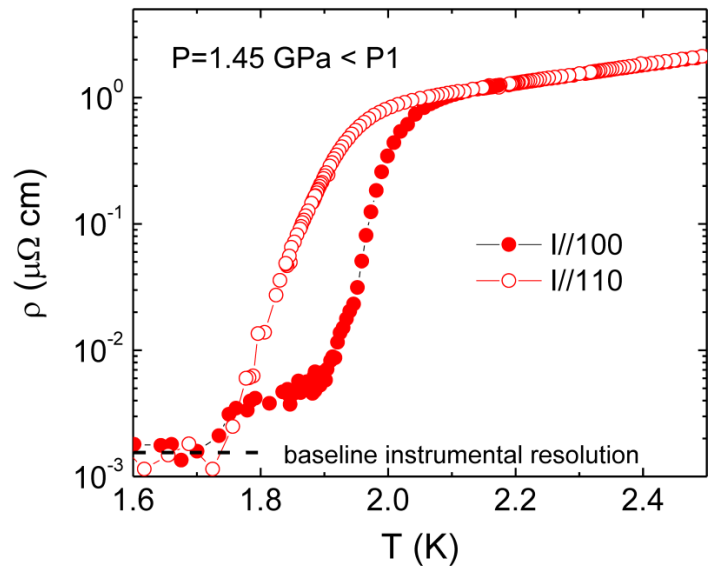

Figure 4

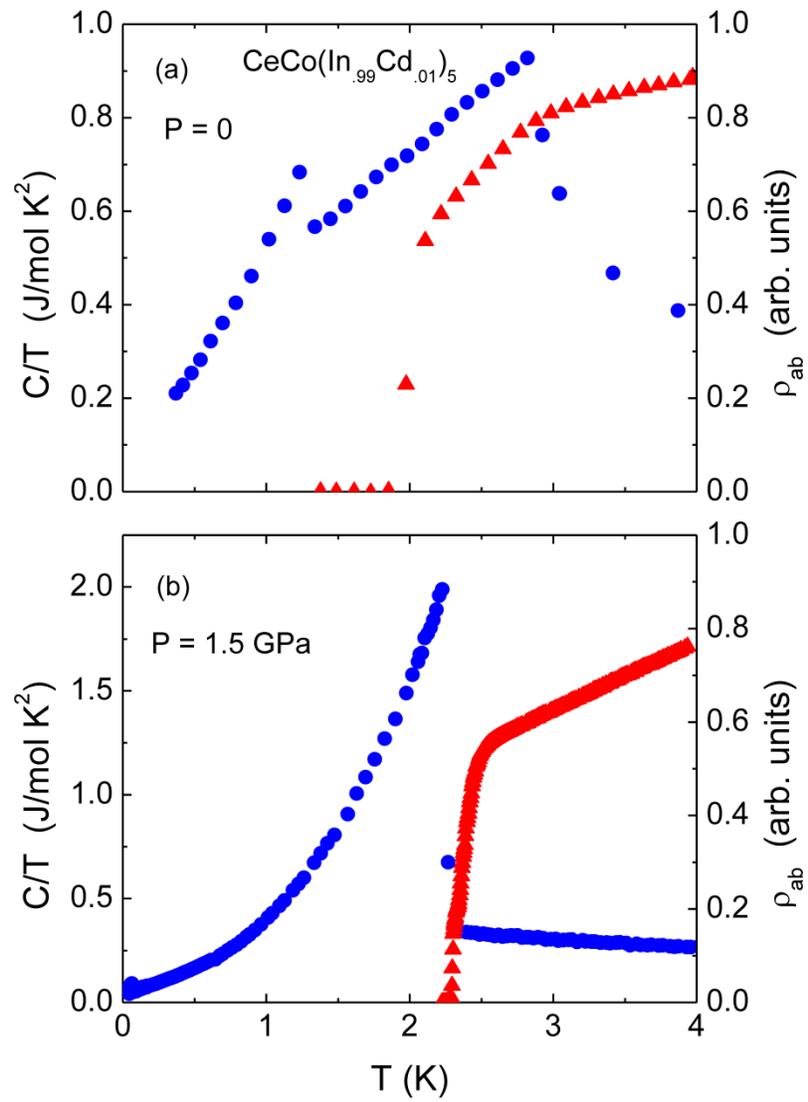

Figure 5

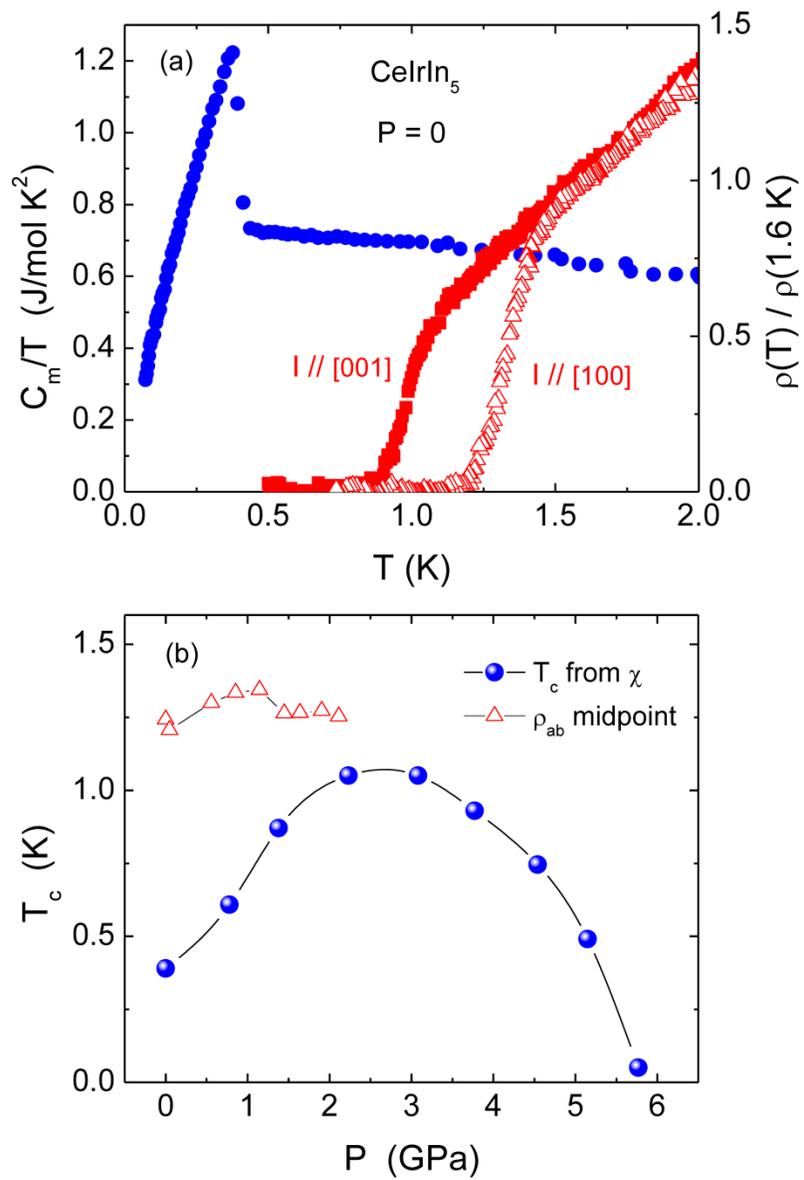

Figure 6

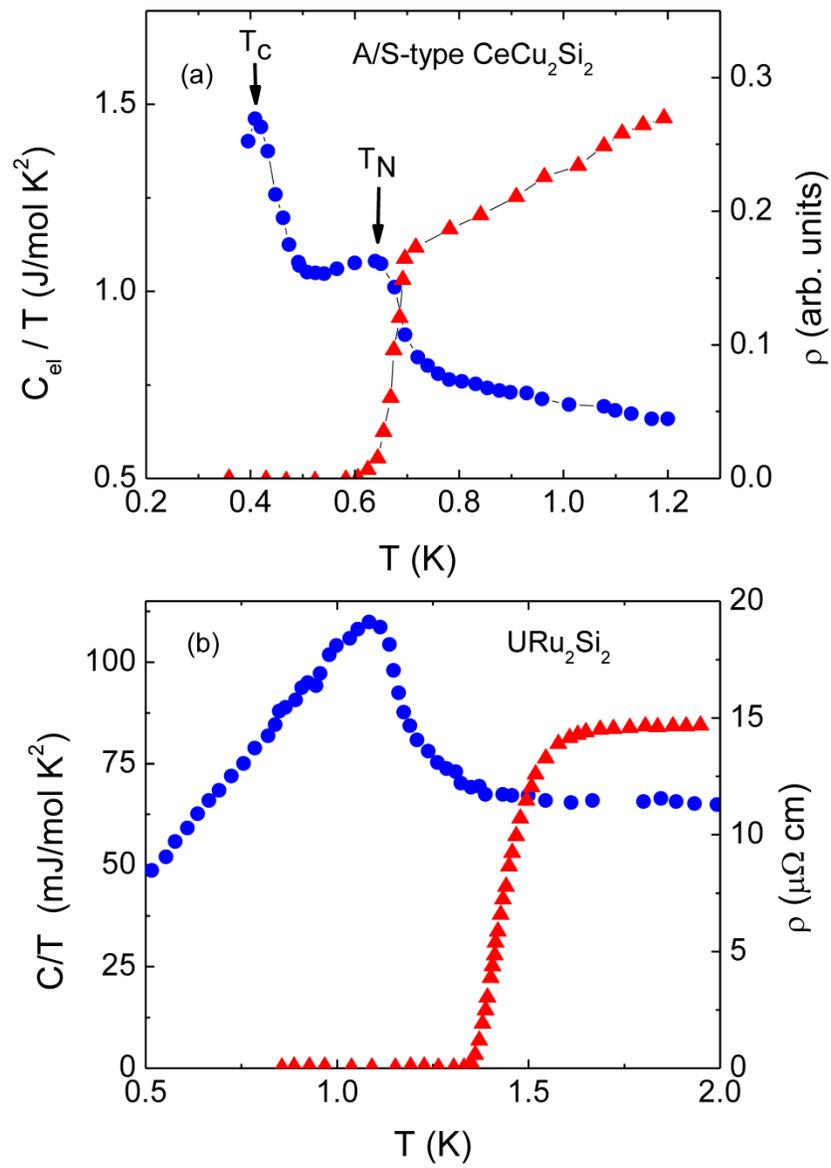

Figure 7